\documentclass[11pt,a4paper]{article}
\usepackage{jcappub}
\usepackage{graphicx}  
\usepackage{dcolumn}   
\usepackage{bm}        
\usepackage{amssymb}   
\usepackage[latin1]{inputenc}
\usepackage{amsmath}
\usepackage{amsfonts}
\usepackage{amssymb}
\usepackage{setspace}
\usepackage{mathtools}
\usepackage{pstricks}
\usepackage{color}

\providecommand{\f}[2]{\frac{{#1}}{{#2}}}

\newcommand{\ee}[1]{\begin{equation}#1\end{equation}}
\newcommand{\ea}[1]{\begin{align}#1\end{align}}

\newcommand{\g}{\gamma}

\title{Decoherence Can Relax Cosmic Acceleration: an Example}

\author[a]{Tommi Markkanen}
\affiliation[a]{Department of Physics, King's College London, Strand, London WC2R 2LS, UK}

\abstract{We investigate back reaction in de Sitter space in an approach where only states that are observationally accessible are included in the density matrix. Using the Bunch-Davies vacuum as the initial condition we find for a conformal scalar field and a cosmological constant that tracing over the unobservable states beyond the cosmological horizon leads to a thermal spectrum of particles and that such a configuration is unstable under semi-classical back reaction. It is concluded that this prescription results in an instability of de Sitter space with a gradually increasing horizon size.}

\emailAdd{tommi.markkanen@kcl.ac.uk}
\begin{document}

\maketitle
\section{Introduction}
Black hole evaporation is arguably the most profound example of the back reaction of quantized matter onto space-time geometry \cite{Hawking:1974sw,Hawking:1974sw2}. For its canonical derivation one uses the semi-classical approach where matter is quantized but space-time is kept classical \cite{Birrell:1982ix}. A crucial step in the closely related information loss calculation is the realization that all degrees of freedom beyond a space-time horizon should be neglected, or traced over, as they are in principle inaccessible to a local observer \cite{Hawking:1976ra}. An analogous step is often implemented when investigating the generation of a thermal spectrum of particles due to accelerated motion, or the Unruh effect \cite{Unruh:1976db,Unruh:1976db2}.

Space-times with horizons, such as de Sitter space, manifest particle creation and admit a thermal description \cite{Gibbons:1977mu}. For an observer at rest with the static coordinate system particle creation gives de Sitter space the characteristics of a "hot tin can" \cite{Dyson:2002pf,Bousso}. Using thermodynamic arguments it was concluded in \cite{Gibbons:1977mu} that de Sitter space is stable. However, in \cite{Padmanabhan:2002ji,Paddy} it was argued, also via  thermodynamics, that de Sitter space evaporates like a black hole. 

The current Universe is expanding at an accelerating rate consistent with the presence of vacuum energy \cite{Betoule:2014frx}, which in a classical approximation can be described as the cosmological constant term in the Einstein-Hilbert Lagrangian. Eventually vacuum energy will lead to the exponentially expanding de Sitter solution, which classically persists forever. 

Here we examine the quantum back reaction in de Sitter space in the semi-classical approach where the expectation value of the energy-momentum tensor is used as the source in the Friedmann equations. Our system will consist of classical vacuum energy and a conformally coupled quantized scalar field for an observer approximately at rest with the flat expanding Friedmann--Lema\^{i}tre--Robertson--Walker (FLRW) coordinate system. Our main focus is investigating the fate of the Universe with a very small Hubble rate $H$ in accord with current observations. At this limit the semi-classical approach is expected to be valid \cite{Birrell:1982ix}. The stability and back reaction in de Sitter space have been extensively studied in the past \cite{Dyson:2002pf,1,2,3,4,5,6,7,8,9,10,11,12,13,14,15,4a,4b,markrajan,Paddy,me}, see \cite{me} for more references. 

Recently in \cite{me} it was discovered that the back reaction in de Sitter space can be drastically altered if there is any loss of information -- or decoherence in the language of \cite{me} -- in the density matrix with which one calculates the expectation values. The focal point of this article is then to make use of the proposition of \cite{me} to investigate the implications of a particular example of decoherence in de Sitter space: loss of observable access due to the cosmological horizon. Considering only those degrees of freedom that are not hidden by a horizon parallels the arguments that were shown to lead to the information paradox in \cite{Hawking:1976ra}. Furthermore, decoherence of the density matrix is a crucial element in the inflationary paradigm without which the inherently quantum perturbations of the inflaton never classicalize and hence do not result in the observed large scale structure \cite{Polarski:1995jg,Kiefer:1998qe}.

We will use the (+,+,+) conventions of \cite{Misner:1974qy} with $\hbar\equiv c\equiv k_B\equiv 1$.
\section{Semi-classical back reaction with a coarse grained density matrix}
\label{sec;2}
The flat FLRW line element $ds^2=-d\tau^2+a^2(\tau)d\mathbf{x}^2$ with $a(\tau)\equiv a$ describes a homogeneous and isotropic Universe and leads from the semi-classical Einstein equations to the (semi-classical) Friedmann equations
\ea{
\begin{cases}\phantom{-(}3H^2M_{\rm pl}^2&= \rho_m+\rho_\Lambda\\ -(3H^2+2\dot{H})M_{\rm pl}^2 &= p_m+p_\Lambda \end{cases}\,,\label{eq:e}}
with $M_{\rm pl}\equiv (8\pi G)^{-1/2}$. The vacuum energy $\rho_\Lambda=-p_\Lambda$  is assumed classical and the  $\rho_m$ and $p_m$ are the renormalized expectation values of the quantized energy- and pressure densities for the conformal scalar field, calculated from the energy-momentum, $\rho_m\equiv\langle\hat{T}_{\tau\tau}\rangle-\delta T_{\tau\tau}$ and $p_m\equiv(\langle\hat{T}_{ii}\rangle-\delta T_{ii})/a^2$, and where $\delta T_{\mu\nu}$ contains the counter terms. The matter action in $n$-dimensions reads $S_m=-\int d^nx\sqrt{-g}\mathcal{L}_m$ with
\ee{2\mathcal{L}_m=(\nabla_\mu \phi)^2+\xi R\phi^2\,;\quad4\xi=(n-2)/(n-1)\label{eq:act1}\,.}
From (\ref{eq:e}) we get our main dynamical relation
\ee{-2\dot{H}M_{\rm pl}^2=\rho_m+p_m\,,\label{eq:dyn}} for analysing the back reaction. Equation (\ref{eq:dyn}) shows that states for which in de Sitter space the sum of the energy- and pressure densities is non-zero imply a deviation from de Sitter due to back reaction, as they result in $\dot{H}\neq0$. 

Following the prescription of \cite{me} we will calculate the expectation values of the energy-momentum tensor used in the Friedmann equations by using a coarse grained density matrix where the unobservable degrees of freedom have been neglected, which we emphasize, is not the usual approach implemented in semi-classical gravity \cite{Birrell:1982ix}. In fact, de Sitter invariance of the Bunch-Davies vacuum \cite{BD,BD2}, which will be our initial condition, demands that the expectation value of the energy-momentum tensor behaves as a cosmological constant and leads to no back reaction if no states of the initial condition are traced over \cite{me}. However, in a realistic scenario a quantum system always exhibits some loss of information/decoherence which affects the results one obtains for the quantum expectation values. In particular, often symmetries are effectively broken due to the limited access of a local observer to the information contained in the initial state \cite{Giulini:1996nw}, which a priori seems to potentially imply important consequences in de Sitter space as it possesses a high degree of symmetry. According to the decoherece program suppression of the quantum interference terms due to de-localization of information into an unobservable sector is in fact required in order for the quantum-to-classical transition to take place. For these reasons it is physically motivated to explore the implications of coarse graining also in the context of gravitational backreaction, see \cite{me} for more discussion and references. 

To understand how one may neglect a set of states in a given initial condition let us first suppose a system initialized to the pure state $\vert \Psi\rangle$. In some orthonormal basis $\vert n\rangle$, we can then express the initial density matrix $\hat{\rho}$ as
\ee{\hat{\rho}=\vert\Psi\rangle\langle\Psi\vert=\sum_n\sum_m c(n)c^*(m)\vert n\rangle\langle m\vert\,;\qquad c(n)\equiv\langle n \vert \Psi\rangle
\,.}
When the system decoheres and loses information the off-diagonal terms become supressed, which in the $\vert n \rangle$ basis we can express as
\ee{\hat{\rho}\quad
\longrightarrow\quad\hat{\rho}^\text{\tiny D}=\sum_n \vert c(n)\vert^2\vert n\rangle\langle n\vert\,,\label{eq:dec0}}
where following \cite{me} the superscript "D" stands for decohered. After decoherence the expectation values can be obtained in the usual manner as

\ee{\langle\hat{\mathcal{O}}\rangle\equiv{\rm Tr}\big\{\hat{\mathcal{O}}\hat{\rho}^\text{\tiny D}\big\}\,.} 

We assumed to initially have a pure state $\hat{\rho}^2=\hat{\rho}$, but for $\hat{\rho}^\text{\tiny D}$ one usually has \ee{(\hat{\rho}^\text{\tiny D})^2\neq\hat{\rho}^\text{\tiny D}\quad\Leftrightarrow\quad0\leq\vert c(n) \vert^2<1\,,}
showing that generically when a system decoheres, its entropy increases since the state is no longer pure but has become mixed. From this one may see how an instability might arise in this prescription: an observer at rest with the FLRW coordinates will perceive a horizon at the physical distance $1/H$. If we then neglect or coarse grain over the unobservable states beyond the horizon it is expected that the entropy of the system is increased in the process. This implies that the system is no longer in its vacuum since the vacuum is a pure state with no entropy. If this leads to the generation of a fluid component behaving as a particle density, it suggests $\dot{H}<0$ since particles do not generally have negative pressure. 

In the FLRW coordinates if $\rho_m>0$, the only stable configurations are ones where 
$\rho_m$ does not consist of particles but of a fluid with negative pressure. For the conformal theory with the Lagrangian (\ref{eq:act1}) the stable configurations are even more constrained: conformal invariance leads classically to a traceless energy-momentum tensor for $\phi$ giving $T_{ii}/a^2=T_{\tau\tau}/3$, and hence any non-zero energy-density gives $\dot{H}\neq 0$. This assertion might be challenged on the basis that the trace does not vanish after quantization due to the conformal anomaly \cite{Capper:1974ic}. But the conformal anomaly results from the counter terms $\delta T_{\mu\nu}$ and it is straightforward to check its effect, for example, by using the results of \cite{me}. What one finds is that the counter terms and hence the conformal anomaly drop out from the right-hand side of (\ref{eq:dyn}) since $\delta T_{\tau\tau}=-\delta T_{ii}/a^2$. This is a consequence of de Sitter symmetry of the $\delta T_{\mu\nu}$, which holds if they are the result of local terms in the Einstein-Hilbert action, as can be checked with the formulae of \cite{Markkanen:2013nwa}. Thus the right-hand side of (\ref{eq:dyn}) becomes $(4/3)\rho_m^S$ as naively deduced from the classically vanishing trace, but where $\rho_m^S$ contains only state dependent terms\footnote{For a non-covariant regularization such as a cut-off the divergences remain and need to be subtracted by hand. In dimensional regularization this is not needed \cite{markrajan}.}.

The above considerations allow us to formulate four conditions which cannot be satisfied simultaneously in a model with vacuum energy and a scalar field: (1) a conformally coupled theory. (2) a FLRW line element. (3) $\rho^S_m\neq0$. (4) $\dot{H}=0$. As we already discussed, after coarse graining one expects non-zero entropy and $\rho_m^S>0$ implying  $\dot{H}\neq 0$. 

Next we investigate the specific example of coarse graning over the states beyond the cosmological horizon in the Bunch-Davies vacuum.
\section{Tracing over unobservable states in the Bunch-Davies vacuum}


\begin{figure}
\begin{center}
\includegraphics[width=0.3\textwidth,angle=90,origin=c,trim={0.3cm 0cm 1.88cm 0cm},clip]{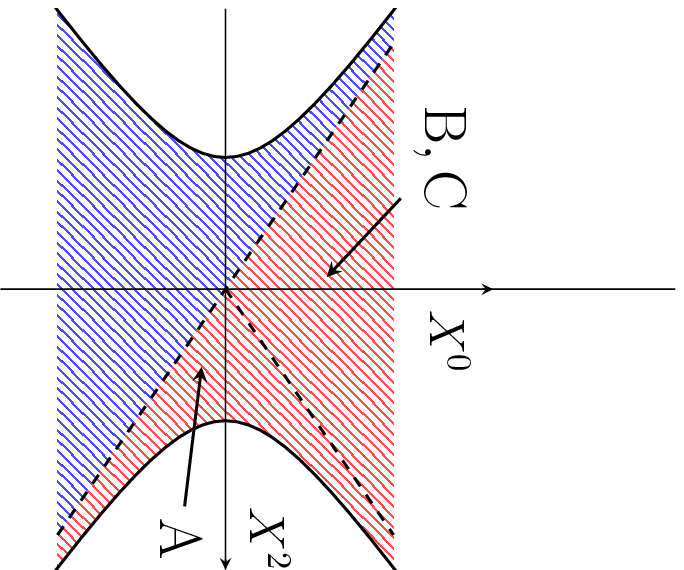}
\end{center}
\caption{Projection of the de Sitter manifold on the $(X^2,X^0)$-plane, where the red patch can be covered by the FLRW coordinates (\ref{eq:FLRW0}), or the $A,B$ and $C$ coordinates with (\ref{eq:staA0}) and (\ref{eq:staB}).  \label{fig:hyper}} 
\end{figure}
The $n$-dimensional de Sitter space is described by an $n$-dimensional hyperboloid embedded in $n+1$ dimensional flat space and in two dimensions is determined by the surface $-(X^0)^2+(X^1)^2+(X^2)^2=H^{-2}$ and $ds^2=-(dX^0)^2+(dX^1)^2+(dX^2)^2$. Its projection on the $(X^2,X^0)$-plane is depicted in Fig \ref{fig:hyper}. 

For simplicity we consider only the 2-dimensional case for which the FLRW coordinate system can be parametrized with
\ea{\begin{cases}X^0&=H^{-1}\sinh(H\tau)+(H/2)Z^2e^{H\tau}\,;\quad X^1=Z e^{H\tau}\\
X^2&=H^{-1}\cosh(H\tau)-(H/2)Z^2e^{H\tau}\end{cases}\,,\label{eq:FLRW0}}
where $\tau \in [-\infty,\infty]$ and $Z\in [-\infty,\infty]$, with 
\ee{ds^2=-d\tau^2+e^{2H\tau}dZ^2\,,\label{eq:FLRW}}
and is denoted with red colour in Fig. \ref{fig:hyper}\,. This highlights the fact that the complete de Sitter manifold contains an expanding patch and a contracting patch, denoted with red and blue respectively in Fig. \ref{fig:hyper}\,. As we currently have knowledge of only an expanding space we wish to obtain a result that is independent of any possible contracting phases of the metric. Hence in this calculation we only consider the patch of the de Sitter manifold that is described by the expanding FLRW coordinates.
 
The expanding FLRW coordinates are the standard choice for inflationary cosmology. What also is standard is choosing the Bunch-Davies (BD) vacuum \cite{BD} as the state for the inflationary power spectrum. This can be motivated by the fact that, given certain assumptions of the short distance behaviour \cite{Allen}, this state is an equilibrium state approached by all initial conditions \cite{markrajan}. For these reasons the BD vacuum is also the suitable one for describing the de Sitter phase of the late time Universe. 

Since in our prescription we only include states accessible to a local observer in the back reaction, we cannot simply take the expectation value of the energy-momentum tensor in the BD vacuum as the source in the Friedmann equation (\ref{eq:e}): a local observer only sees the patch inside the horizon, which is described with the static coordinates
\ea{\begin{cases}X^0&=(H^{-2}-z_A^2)^{1/2}\sinh(H t_A)\,;\quad X^1=z_A\\
X^2&=(H^{-2}-z_A^2)^{1/2}\cosh(H t_A)\end{cases}\,,\label{eq:staA0}}
where $t_A \in [-\infty,\infty]$ and $z_A\in [-1/H,1/H]$, with
\ee{ds^2=-\big[1-(Hz_A)^2\big]dt_A^2+\big[1-(Hz_A)^2\big]^{-1}dz_A^2\,,\label{eq:staA}}
where the horizons at $z_A=\pm H^{-1}$ are now explicit. This only covers one-quarter of the entire de Sitter manifold, the region $A$ in Fig. \ref{fig:hyper}\,. For the remaining part of the expanding FLRW patch we can use
\ea{\begin{cases}X^0&=(z_B^2-H^{-2})^{1/2}\cosh(H t_B)\,;\quad X^1=z_B\\
X^2&=(z_B^2-H^{-2})^{1/2}\sinh(H t_B)\end{cases}\,,\label{eq:staB}}
with $t_B \in [-\infty,\infty]$ and $z_B\in [1/H,\infty]$ for the region $B$ which lies behind the horizon $z_A=1/H$ and for the region $C$ behind $z_A=-1/H$ we can use coordinates as in (\ref{eq:staB}) but with $z_C\in [-\infty,-1/H]$. Both lead to the form (\ref{eq:staA}) for the line element. In the projection of Fig. \ref{fig:hyper}\,, the regions $B$ and $C$ appear on top of each other.

In order to obtain a coarse grained density matrix including only the observable states, we must trace over the regions $B$ and $C$ in our initial condition, the BD vacuum. Symbolically, we will write the resulting density matrix as
\ee{\hat{\rho}^\text{\tiny D}\equiv\text{Tr}_{BC}\{|0\rangle\langle0|\}\,,}
where $\hat{\rho}^\text{\tiny D}$ denotes the coarse grained density matrix as in the previous section, $|0\rangle$ the BD vacuum and $\text{Tr}_{BC}$ the trace over the hidden states in the $B$ and $C$ regions.

\begin{figure}
\begin{center}
\includegraphics[width=0.4\textwidth,trim={0 1.3cm 0 0},clip]{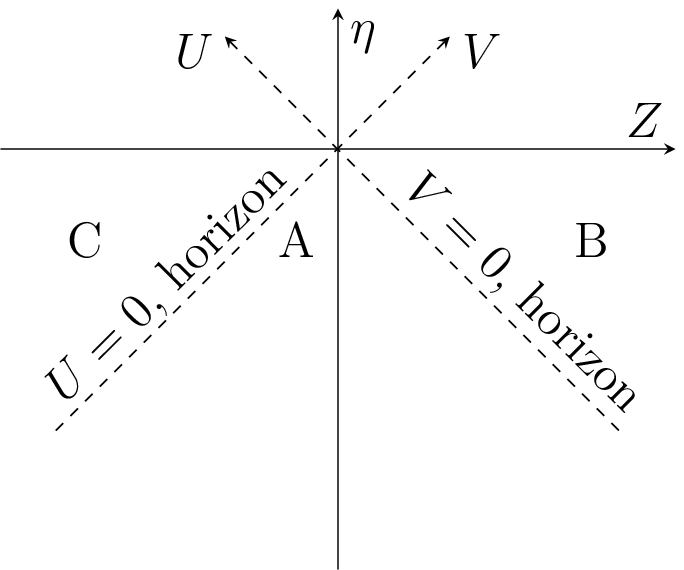}
\end{center}
\caption{Relation of the regions $A,B$ and $C$ to the horizons in $(Z,\eta)$- and $(V,U)$-coodinates. \label{fig:horizons}} 
\end{figure}

Next we need relations between the various coordinate systems. It proves convenient to first transform
\ee{\eta=-H^{-1}e^{-H\tau}\quad\text{and}\quad z^*=(2H)^{-1}\log\bigg\vert\f{1+zH}{1-zH}\bigg\vert\,,}
for the FLRW and $A,B,C$ coordinate systems, respectively. Here $\eta\in [-\infty,0]$ is the usual conformal time familiar from the cosmological context, and  $z^*_A\in [-\infty,\infty]$, $z^*_B\in [0,\infty]$ and $z^*_C\in [-\infty,0]$ are the tortoise coordinates used frequently in black hole physics. A final convenient transformation is introducing the standard light-cone coordinates valid for $z_A$, $z_B$ and $z_C$
\ea{V&=\eta+Z=+e^{-H\tau}\big(z-H^{-1}\big)\label{eq:LC0}\,,\\
U&=\eta-Z=-e^{-H\tau}\big(z+H^{-1}\big)\,.\label{eq:LC}}
These coordinates provide an important illustration of the relation of the horizons at $V=0$ and $U=0$ to the regions $A,B$ and $C$, as shown in Fig. \ref{fig:horizons}\,.
Using $U$ and $V$ we can express the FLRW coordinates with the coordinates of the regions $A,B$ and $C$
\ea{V&\overset{A}{=}-H^{-1}e^{-Hv_{A}}\overset{B}{=}+H^{-1}e^{-Hv_B}\overset{C}{=}-H^{-1}e^{-Hv_C}\,,\nonumber\\
U&\overset{A}{=}-H^{-1}e^{-Hu_A}\overset{B}{=}-H^{-1}e^{-Hu_B}\overset{C}{=}+H^{-1}e^{-Hu_C}\,,\label{eq:trans}}
where the oversets above the equalities are to be understood to indicate in which region the relation is valid and we also used $v=t+z^*$ and $u=t-z^*$. The relations (\ref{eq:trans}) are very similar to the transformations between a stationary and a uniformly accelerated frame in Minkowski space \cite{Birrell:1982ix,Crispino}. For this reason, much of the calculation can be performed with the standard steps needed for the Unruh effect \cite{Birrell:1982ix,Crispino}, see \cite{Carroll:2004st} for a particularly clear derivation.  The sign changes in (\ref{eq:trans}) when a horizon is crossed are crucially important and in a way, the mathematical reason behind the result that follows.

The equation of motion for the two-dimensional conformal quantum field with the action (\ref{eq:act1}) in the $(\eta,Z)$ coordinates is the same as in flat space with the solution
\ea{&\hat{\phi}(V,U) =\int_0^\infty\f{d\omega}{\sqrt{4\pi\omega}}\Big[g(V)\hat{a}^{\phantom{\dagger}}_\omega +g^*(V)\hat{a}^{\dagger}_\omega+h(U)\hat{a}^{\phantom{\dagger}}_{-\omega}+h^*(U)\hat{a}^{\dagger}_{-\omega}\Big]
\equiv\hat{\phi}(V)+\hat{\phi}(U)\,,\label{eq:mode}}
the modes $g(V)=e^{-iV\omega}$ and $h(U)=e^{-iU\omega}$, and the $[\hat{a}_{\omega^{\phantom{.}}}^{\phantom{\dagger}},\hat{a}_{\omega'}^\dagger]=\delta(\omega-\omega')$ convention for the commutators.
The vacuum state defined by the annihilation operators $\hat{a}_{\omega}$ and $\hat{a}_{-\omega}$
is the 2-dimensional BD vacuum.

In order to obtain solutions for the quantum field in the regions $A,B$ and $C$ we must find the appropriate future oriented and time-like vectors with which to define inner products. In region $A$ with the help of (\ref{eq:trans}) we can use $\partial_{t_A}=-H(\eta\partial_\eta+Z\partial_Z)$, since $-\eta>0$ and $\eta^2>Z^2$ in this region. Following the same procedure, we find that for regions $B$ and $C$ we can use $-\partial_{z^*_B}$ and $\partial_{z^*_C}$, respectively. The scalar products for the regions $A,B$ and $C$ are then
\ee{i(y,\bar{y})_A=\int dz^*_A\,y\overset{\leftrightarrow}{\partial}_{t_A}\bar{y}^*\,;\qquad i(y,\bar{y})_C=\int dt_C\,y\overset{\leftrightarrow}{\partial}_{z^*_C}\bar{y}^*\,;\qquad i(y,\bar{y})_B=-\int dt_B\,y\overset{\leftrightarrow}{\partial}_{z^*_B}\bar{y}^*\,,\label{eq:ip}} where $y$ and $\bar{y}$ are two arbitrary modes and $\overset{\leftrightarrow}{\partial}=\overset{\rightarrow}{\partial}-\overset{\leftarrow}{\partial}$. With (\ref{eq:ip}) one may define properly normalized modes, which when expressed as an integral over only positive values of $\omega$ as in (\ref{eq:mode}) read\footnote{The spatial part of the mode is chosen to have opposite signs to the temporal. E.g in reg. $B$, $z^*_B$ describes time and $\hat{\phi}=\int{dk}/{\sqrt{4\pi|k|}}\big(e^{i(z^*_B|k|-t_B k)}\hat{a}^B_{-k}+ {\rm H.C.}\big)$, with a conventional '$-$' sign in $\hat{a}^B_{-k}$. Positive frequency is defined w. r. t. $-\partial_{z^*_B}$, and finally $\int =\int_0^\infty+\int_{-\infty}^0$ gives (\ref{eq:modes}).}
\ea{g^A(v_A)&=e^{-iv_A\omega}\,;\qquad
h^A(u_A)=e^{-iu_A\omega}\label{eq:modes0}\,,\\ \label{eq:modes}g^B(v_B)&=e^{+iv_B\omega}\,;\qquad
h^B(u_B)=e^{-iu_B\omega}\,,\\g^C(v_C)&=e^{-iv_C\omega}\,;\qquad
h^C(u_C)=e^{+iu_C\omega}\,,\label{eq:modes1}} 
where the sign flips again reflect the horizon structure.

The main objective of this calculation is to express the BD vacuum as a product state defined by the operators of the various regions, so that we can precisely identify the states to be traced over. Since $[\hat{a}^{\phantom{\dagger}}_\omega,\hat{a}^{{\dagger}}_{-\omega}]=0$ \footnote{We neglect the zero modes.}, $\hat{\phi}(V)$ commutes with $\hat{\phi}(U)$ and further since $V=V(v)$ and $U=U(u)$ from (\ref{eq:trans}), the $V$- and $U$-sectors can be analysed separately. Formally, we can thus write $|0\rangle=|0_V\rangle\otimes|0_U\rangle$ and $\hat{\rho}=\hat{\rho}^V\otimes\hat{\rho}^U$.

Considering only the regions $A$ and $B$ and $\hat{\rho}=\hat{\rho}^V$, we write the field with the $A$ and $B$ modes from (\ref{eq:modes0}--\ref{eq:modes})
\ee{\hat{\phi}(V)=\int^\infty_0\f{d\omega}{\sqrt{4\pi\omega}}\Big[g^A(v_A)\hat{a}^{A}_\omega+g^B(v_B)\hat{a}^{B}_\omega+\text{H.C.}\Big]\,,\label{eq:lincomb}} 
where the $A$ mode vanishes in region $B$ and vice versa for the $B$ mode. The operators $\hat{a}^{A}_\omega$ and $\hat{a}^{B}_\omega$ are the annihilation operators as defined by $g^A(v_A)$ and $g^B(v_B)$, and in general are inequivalent to $\hat{a}_\omega$ as defined by (\ref{eq:mode}).

Much like in \cite{Unruh:1976db} we now study the analyticity properties of the modes. With (\ref{eq:trans}) we can write (\ref{eq:modes0}--\ref{eq:modes}) as
\ee{g^A(v_A)=e^{i(\omega/H)\log\left(-HV\right)}\,;\quad \label{eq:nmode1} g^B(v_B)=e^{-i(\omega/H)\log\left(HV\right)}\,.}
Since the mode $g(V)$ in (\ref{eq:mode}) is analytic in the lower complex $V$-plane, the $g^A(v_A)$ and $g^B(v_B)$ cannot both be linear combinations of $g(V)$-modes due to the discontinuity at the horizon $V=0$. However, if we choose the complex logarithm to have a branch cut on the negative imaginary axis, the combinations 
\ea{f^{(1)}&=Ce^{+\pi\omega/(2H)}\Big[g^A(v_A)+e^{-\pi\omega/H}g^{B\,*}(v_B)\Big]\,,\label{eq:f}\\
f^{(2)}&=Ce^{-\pi\omega/(2H)}\Big[g^{A\,*}(v_A)+e^{+\pi\omega/H}g^{B}(v_B)\Big]\,,\label{eq:f1}}
where $C$ is a normalization constant, are analytic for $\Im[V]<0$ and hence are expressable as a linear combination of the $g(V)$-modes with the same vacuum \cite{Birrell:1982ix}. Fixing $|C|^{-2}=2\sinh(\pi\omega/H)$ we can express (\ref{eq:lincomb}) also as
\ee{\hat{\phi}(V)=\int^\infty_0\f{d\omega}{\sqrt{4\pi\omega}}\Big[f^{(1)}\hat{d}^{(1)}_\omega+f^{(2)}\hat{d}^{(2)}_\omega+\text{H.C.}\Big]\,,
\label{eq:lincomb2}}
where $\hat{d}^{(1)}_\omega|0\rangle=\hat{d}^{(2)}_\omega|0\rangle=0$. From (\ref{eq:lincomb}) and (\ref{eq:lincomb2}) one may find relation between the two sets $(\hat{d}^{(1)}_\omega,\hat{d}^{(2)}_\omega)$ and $(\hat{a}^{A}_\omega,\hat{a}^{B}_\omega)$, and write $\vert 0_V\rangle$ as an entangled combination of the states as defined by the $A$ and $B$ modes \cite{Fiola:1994ir,Crispino} 
\ee{\vert 0_V\rangle_\omega =\sqrt{1-\gamma^2}\sum_{n_{\omega}=0}^\infty\gamma^{n_\omega}\vert n_\omega,A\rangle\otimes\vert n_\omega,B\rangle\,;~~ \gamma\equiv e^{-\f{\pi\omega}{H}}\,,}
where $\vert n_\omega,A\rangle$ stands for a state with $n_\omega$ particles with the momentum $\omega$ as defined by the $A$ modes and similarly for $B$ and we denote the $\omega$'th oscillator contribution of $\vert 0_V\rangle$ as $\vert 0_V\rangle_\omega$. Tracing over the $\vert n_\omega,B\rangle$ states leads to the decohered density matrix as the product, $\hat{\rho}^V\equiv\Pi^{\infty}_{\omega=0}\hat{\rho}^V_\omega$, with
\ee{\big(\hat{\rho}_\omega^V\big)^\text{\tiny D}=(1-\gamma^2)\sum_{n_{\omega}=0}^\infty\gamma^{2n_\omega}\vert n_\omega,A\rangle\langle A,n_\omega\vert\,,\label{eq:therm}}
i.e. a thermal density matrix with the Gibbons-Hawking de Sitter temperature $T=H/(2\pi)$.

Since (\ref{eq:therm}) includes particle creation from the region $B$ only but the full BD state extends also to region $C$, one in principle could have included it in the above analysis. But from  (\ref{eq:modes0}), (\ref{eq:modes1}) and (\ref{eq:trans}) we see that $g^C(v_C)$ has the same functional form as $g^A(v_A)$ and no mixing of the positive and negative frequency modes is needed for an analytic behaviour accross $U=0$ and hence there is  no additional particle creation. When studying only the $V$-sector this is a perfectly natural result, since the horizon at $U=0$ is not felt by a function only dependent on $V$.

Particle creation in the $U$-sector comes via a similar derivation. The same reasoning we used for the $V$-sector now implies for the $U$-sector that particles are only produced by the horizon $U=0$ and hence entanglement between the $A$ and $C$ modes. This, and the resulting thermal density matrix one may verify by a straightforward calculation. Note that from (\ref{eq:mode}) we see that $\hat{\phi}(V)$ contains only $\hat{a}^{\phantom{\dagger}}_\omega$ and $\hat{a}^{\dagger}_\omega$ where as $\hat{\phi}(U)$ only $\hat{a}^{\phantom{\dagger}}_{-\omega}$ and $\hat{a}^{\dagger}_{-\omega}$, so overall momentum conservation is satisfied. 

\subsection{The energy-momentum tensor}
The initial BD vacuum after coarse graining is precisely thermal for the modes in the observable $A$ region as we can see from (\ref{eq:therm}). We can use the decohered density matrix to calculate the expectation value of the energy-momentum tensor by writing $\hat{\phi}$ in terms of $(g^A(v_A),\hat{a}^A_\omega)$ and $(h^A(u_A),\hat{a}^A_{-\omega})$ to get
\ee{\langle\hat{T}_{vv}\rangle=\langle\hat{T}_{uu}\rangle=\int_0^\infty\f{d\omega}{2\pi}\bigg[\f{\omega}{2}+\f{\omega}{e^{2\pi\omega/H}-1}\bigg]\,;~~~\langle\hat{T}_{uv}\rangle=0\,,\label{eq:tuu}}
where for simplicity we have dropped the '$A$' labels. For small vacuum energy, which is our main focus, we can take the limit of a large horizon radius, $|z|H\ll1$, and make use of an expansion in $zH$. From (\ref{eq:tuu}) it follows that at zeroth order the energy-momentum is homogeneous and isotropic. Since this is true also of the generated divergences, the result can be renormalized with the adiabatic procedure of \cite{Bunch} which is equivalent to a redefinition of the parameters in the action \cite{Markkanen:2013nwa}. The two-dimensional calculation can be found in \cite{Bunch2} which gives coinciding results with \cite{Davies:1976hi}. Covariant conservation of the counter terms allows one to iteratively solve the linear order corrections to $\delta T_{\mu\nu}$,  with which (\ref{eq:tuu}) give in the FLRW coordinates 
\ea{T^R_{\eta\eta}&=e^{2H\tau}\bigg[\int_{-\infty}^\infty\f{dk}{2\pi}\f{|k|}{e^{2\pi|k|/H}-1}-\f{H^2}{24\pi}\bigg]\,,\label{eq:end1}\\ T^R_{ZZ}&= T^R_{\eta\eta}+\f{e^{2H\tau} H^2}{12\pi}\,;\quad T^R_{\eta Z}=z H\big(T^R_{ZZ}+T^R_{\eta\eta}\big)\label{eq:end2}}
where ${T}^R_{\mu\nu}\equiv\langle \hat{T}_{\mu\nu}\rangle-\delta T_{\mu\nu}$ and we have neglected terms of $\mathcal{O}((zH)^2)$. This energy-momentum is covariantly conserved and gives rise to the standard conformal anomaly \cite{Birrell:1982ix}. The presence of the flux term $T^R_{\eta Z}$ is expected from energy conservation: to have a thermal energy density with constant temperature in an expanding space, a flux is required to counteract the dilution from expansion. Close to the origin, a large region for small $H$, the energy-momentum is homogeneous and isotropic since the fluxes from both sides are equal. If  $\rho_m^S$ is the state dependent thermal contribution given by the integral in (\ref{eq:end1}), the sum of the pressure and energy components is $\rho_m+p_m=2\rho_m^S$ as naively implied by the classically vanishing trace, analogously to what was discussed in the four dimensional case at the end of section \ref{sec;2}.
\section{Conclusions}
We can conclude that in two dimensions at the limit of a large horizon radius when only including the observable states in the expectation value of energy-momentum tensor for a conformal scalar field the system is homogeneous and isotropic with a non-zero energy density $\rho^S_m$ that is precisely thermal with the temperature $H/(2\pi)$.
If we assume that this persists also in 4 dimensions it implies an instability of de Sitter space in this prescription. For small $\rho_\Lambda$ it is expected that back reaction is weak and takes a long time to cumulate, so to a good approximation we can use the results calculated on fixed background on the right-hand side of (\ref{eq:dyn}) and solve $H$ self-consistently, much like what one does in the standard black hole calculation \cite{Hawking:1974sw}. Using a 4-dimensional thermal energy-density with $T=H/(2\pi)$ as $\rho_m^S$ and $\rho_m+p_m=(4/3)\rho^S_m$ we find
\ee{-\dot{H}M_{\rm pl}^2=\f{H^4}{720\pi^2} \quad\Rightarrow\quad \f{H}{H_0}={\bigg(\f{H_0^3\tau}{240\pi^2 M_{\rm pl}^2}+1 \bigg)^{-1/3}}\,,\label{eq:Hev}}
with $H_0\equiv H(0)$ and where we used $\rho_m^S=\f{\pi^2}{30}T^4$. The result (\ref{eq:Hev}) coincides with \cite{Padmanabhan:2002ji} (section 10.4) and is a special case of the solutions considered in \cite{me}. From (\ref{eq:Hev}) we see that the de Sitter approximation is inconsistent for large time scales. We can estimate the breakdown of the approximation by calculating the half-life $H(\tau_{1/2})=H_0/2$, which gives $\tau_{1/2}\sim M_{\rm pl}^2/H_0^3$ and agrees with the 'break-time' given in \cite{Dvali:2017eba}. Evidently, in our approach quantum back reaction can destabilize the de Sitter solution, but for $H_0\ll M_{\rm pl}$ this will take an enormously long time, more than $10^{100}$-times the age of the Universe for the observed $H_0\sim10^{-42}$GeV.

From a physical point of view, neglecting the unobservable degrees of freedom as implemented here and in \cite{me} seems like a well-motivated prescription. After all, this results in an expectation value for the energy-momentum tensor that corresponds to what a local observer would actually measure. However, this is in sharp contrast to what is generally accepted as the semi-classical approach \cite{Birrell:1982ix}. If one would simply use the un-coarse grained Bunch-Davies vacuum in the calculation the result would strictly imply $\dot{H}=0$, which follows from the de Sitter invariance of the Bunch-Davies vacuum \cite{markrajan}. At its core, this discrepancy is a question of how  precisely one should formulate the semi-classical approach, which does not have an obvious correct answer and perhaps ultimately requires experimental input. 

The gradual increase of the Hubble horizon due to thermal particle creation as implied by (\ref{eq:Hev}) bears many similarities to black hole evaporation and thermodynamics which warrants further investigation. In the effective sense, this result may be interpreted as dissipating vacuum energy which may have implications for the cosmological constant problem, although it is not obvious how precisely such a decay takes place. These issues as well as a proper 4-dimensional calculation will be discussed elsewhere \cite{mar}.

\acknowledgments{TM would like to thank Arttu Rajantie and Malcolm Fairbairn for discussions. The research leading to these results has received funding from the European Research Council under the European Union's Horizon 2020 program (ERC Grant Agreement no.648680).}

\end{document}